\documentclass[12pt]{article}
\usepackage{epsfig}
\usepackage{amsmath}
\usepackage{amssymb}

\usepackage{latexsym, graphicx}
\newcommand{\be}{\begin{equation}}
\newcommand{\ee}{\end{equation}}
\newcommand{\bea}{\begin{eqnarray}}
\newcommand{\eea}{\end{eqnarray}}

\begin{document}

\title{\vbox{\baselineskip15pt
\hfill \hbox{\normalsize
{hep-ph/0610375}} \\
\hfill \hbox{\normalsize {ANL-PR-06-69}} \\
\hfill \hbox{\normalsize {EFI-06-19}} }
{\bf Baryogenesis\\ from an Earlier Phase Transition}}

\author{{Jing Shu$^{a,b}$\thanks{jshu@theory.uchicago.edu}, \
Tim M.P. Tait$^{b}$\thanks{tait@anl.gov}, \
and Carlos E.M. Wagner$^{a,b}$\thanks{cwagner@hep.anl.gov}}\\[0.5cm]
\normalsize{$^{a}$ Kavli Institute for Cosmological Physics, Enrico Fermi Institute and Department of Physics,}\\[-0.3cm]
\normalsize{University of Chicago, 5640 S. Ellis Avenue, Chicago, IL 60637, USA}\\
\normalsize{$^{b}$ HEP Division, Argonne National Laboratory, 9700
Cass Ave., Argonne, IL 60439, USA} }

\maketitle

\begin{abstract}
We explore the possibility that the observed baryon asymmetry of the
universe is the result of an earlier phase transition in which an
extended gauge sector breaks down into the
$SU(3)_C \times SU(2)_L \times U(1)_Y$
of the Standard Model.  Our proto-typical example
is the Topflavor model, in which there is a separate $SU(2)_1$ for the
third generation from the $SU(2)_2$ felt by the first two generations.
We show that the breakdown of $SU(2)_1 \times SU(2)_2 \rightarrow SU(2)_L$
results in lepton number being asymmetrically distributed through-out the
three families, and provided the SM electroweak phase transition is
{\em not} strongly first order, results in a non-zero baryon number, which
for parameter choices that can be explored at the LHC, may explain the
observed baryon asymmetry.
\end{abstract}

\vfill\eject

\parindent=20pt
\baselineskip=14pt

\section{Introduction}

The origin of the observed baryon asymmetry of the universe
is one of the most
fundamental open questions in particle physics and cosmology.
Numerically, the baryon-to-entropy ratio is
\begin{equation}\label{eta_B} \eta_{B}
\equiv \frac{n_{B}}{s} = 9.2^{+0.6}_{-0.4} \times
10^{-11} \, ,
\end{equation}
where $n_{B} = n_{b} - n_{\bar{b}}$ and $s$ are the
baryon number density and entropy, respectively.  Famously,
Sakharov has shown \cite{Sak67} that in order to generate a
baryon asymmetry in an initially baryon-symmetric universe, there
must be: (1) baryon number ($B$) violation; (2) $C$ and $CP$
violation; and (3) a departure from thermal equilibrium.
These requirements, especially in light of
null results obtained by experimental searches for $B$-violation
(for example, proton decay), are difficult
to realize in the framework of an electroweak (EW) scale
model, prompting attention to high scale mechanisms such as
GUT baryogenesis \cite{Yoshimura,SDLS,Weinberg,EKMST},
leptogenesis \cite{LG86,WBMP} and the Affleck-Dine scenario~\cite{IAMD}.
Such mechanisms can be viable, but are difficult to test experimentally.

However, the Standard Model (SM) already contains a way to reconcile large
baryon-number violation at the EW scale with the lack of experimental evidence
for such interactions at low scales.  As noted by Kuzmin, Rubakov and
Shaposhnikov \cite{Kuzmin:1985mm}, the baryon-number violation
induced by sphaleron transitions between two different $N$-vacua
of the $SU(2)_L$ electroweak interaction are unsuppressed at temperatures
larger than
the EW scale, but are so highly suppressed at low energies as to be
essentially irrelevant.  This idea of electroweak
baryogenesis (EWBG) \cite{BG} is very attractive, and could have been realized
within the SM.
EWBG makes very specific demands of the electroweak symmetry-breaking (EWSB)
sector of the theory, and leads to observable consequences at future colliders.
For example, the need for $B$-violating interactions to be out of equilibrium
at the time of the phase transition requires that the electroweak phase
transition is strongly first order, with
\bea
\frac{\langle \phi(T_c) \rangle}{T_c} & \gtrsim & 1~.
\label{eq:sm1st}
\eea
This in turn puts constraints on the Higgs potential, and demands a small
Higgs quartic (and hence Higgs mass).

Electroweak baryogenesis in the SM is thus ruled out; the LEP-II
bound on the SM Higgs mass~\cite{Barate:2003sz}, $m_h^{SM} \gtrsim
115$~GeV is incompatible with a the requirement that the sphaleron
processes be out of equilibrium at the phase transition,
Eq~(\ref{eq:sm1st}).  While this renders the SM itself unable to
produce the asymmetry, it illustrates the fact that experiments at
the TeV scale are able to directly probe EWBG. Extensions of the
SM which include new physics at the EW scale\footnote{In addition
to new fields, a greatly increased Hubble
expansion\cite{Joyce:1996cp,Joyce:1997fc,Davidson:2000dw,Servant:2001jh}
can re-open some of the Higgs mass parameter space.  Such models
usually require the very fast Hubble expansion to \textsl{only}
happen close to the electroweak scale.} can in fact re-open the
possibility of EW baryogenesis by introducing additional bosonic
fields \cite{Pietroni,ADFM,SJHMGS,Kang:2004pp,Menon:2004wv} as in
the Minimal Supersymmetric Standard Model
(MSSM)~\cite{CJK,CK,CQRVW, MCQCW,CMQSW1,CQSW2,KPSW1,KPSW2} or
fermions strongly coupled to the Higgs
\cite{Carena:2004ha,Provenza:2005nq}. Eventually, future colliders
such as the Large Hardron Collider (LHC) and possibly
International Linear Collider (ILC) will unravel the nature of
EWSB, and should provide a better understanding as to the nature
of the electroweak phase transition (EWPT).  If it proves first
order, this will be a crucial piece of indirect evidence for EWBG.
If it does not, it will raise the interesting question:  Does the
idea of electroweak baryogenesis have to be abandoned?

While the lack of a first order EWPT {\em would} strongly disfavor electroweak
baryogenesis, it may be that the basic picture of baryon-number violating
processes arising from non-perturbative gauge theory dynamics, impotent at
low energies but unsuppressed at the TeV scale, can survive.  Since many
extensions of the SM predict the existence of new non-Abelian gauge symmetries
at the TeV scale, it may be that some theories already addressing unrelated
problems may in fact contain the ingredients necessary for
``electro-weak style'' baryogenesis.  A new gauge group $G$ could break down
somewhere above the EWPT, and generate an asymmetry through its own
strongly first order phase transition.

This is a novel idea, but one which is not entirely
straight-forward to realize in a realistic setting.  Below this
new phase transition, the theory is still $SU(2)_L \times U(1)_Y$
symmetric, and the usual EW sphalerons are unsuppressed.  They
will try to wash-out any generated baryon asymmetry unless $B-L
\neq 0$.  In principle, one could arrange the representations of
the SM matter under the new gauge symmetry $G$ such that $B-L$ is
not preserved due to the nonzero mixed anomaly
$G$-$G$-$U(1)_{B-L}$, but this kind of non-universal assignment of
representations is also somewhat delicate. The requirement that
$G$-$G$-$U(1)_Y$ gauge anomalies cancel will in general require
more chiral fermions charged under $G$ and the SM. Thus, we will
not consider this possibility in detail. Instead, we employ a more
subtle mechanism in which $G$ acts only on the third generation
fermions, and thus $B-L$ is conserved (within a generation). The
breakdown of $G$ can thus be accompanied by production of an
asymmetry among the third family quarks and leptons. Baryon number
will quickly equilibriate among the three families because of the
large quark masses and mixings, but each lepton family number will
remain distinct because of the tiny neutrino masses. Above the
EWPT, the EW sphalerons will result in $B=0$ and $L_1 = L_2 = -L_3
/2 \neq 0$. Provided the EW sphalerons are {\em in} equilibrium
during the EWPT, as the fermion masses turn on from EW
symmetry-breaking, a non-zero (though diluted) baryon number will
result \cite{Kuzmin:1987wn,Dreiner:1992vm}. So in fact this
mechanism {\em requires} that the EWPT {\em not} be strongly first
order.

The outline of this paper is as follows.  In Section~\ref{dilution}, we
review how a baryon asymmetry can be
generated even when $B-L=0$, provided there is a non-zero asymmetry in the
third family lepton number, and how this eventually translates into
a baryon asymmetry after the EWPT.
We apply this mechanism to the
``Topflavor'' model \cite{Chivukula:1995gu}
which is known to contain non-perturbative
interactions which violate baryon- and lepton-number in the third family
\cite{Morrissey:2005uz}.  In fact, we find that it is possible to generate
a baryon asymmetry of the right magnitude. The Topflavor model,
phase transition, $CP$ violating sources,
diffusion equations and calculated baryon number density
are considered in Section~\ref{tf-model}.
Our results show that the right baryon number density can be generated for
parameters that would render this model testable at the LHC.
In Section~\ref{discussion} we present our conclusions.

\section{$B$ from a Family Asymmetric Distribution of $L$}
\label{dilution}

In this section we review the mechanism by which an initial condition
that has $B=L=0$ can nonetheless result in a non-zero baryon number density
provided $L_1 = L_2 = -L_3 / 2 \neq 0$ \cite{Kuzmin:1987wn,Dreiner:1992vm}.
We will show how the specific
example of the Topflavor model can generate these initial conditions through
the non-perturbative dynamics of its phase transition
(in a very similar way to that in which
baryon number is generated in a traditional EWBG
scenario) in Section~\ref{tf-model}.  Thus, for now we assume that
the initial phase transition has generated
$L_3 = B = \Delta$ and $L_1 = L_2 =0$.
The unsuppressed EW sphalerons will rapidly evolve this into a state
with $B=L=0$ but $L_1 = L_2 = -\Delta / 3$ and $L_3 = +2 \Delta / 3$.
We now study what happens as the universe moves through the EWPT, and show
that the non-zero $\tau$ lepton mass will result in $B \sim 10^{-6} \Delta$.
Our discussion closely follows that of Ref.~\cite{Dreiner:1992vm}.

The smallness of observed neutrino masses indicates that lepton-number
violation is out of equilibrium at the EW scale.  Thus, each
lepton flavor has a seperate chemical potential $\mu_i$ with
$i=1,2,3$.  This is in contrast to baryons, because the large quark
masses and mixings keep baryon flavor violation in equilibrium, and thus
the quarks are described by a single chemical potential $\mu$.
We consider the SM matter consisting of three families each of
which consists of two
quarks (an up-type and down-type) with masses $m_{q_i}$, a charged
lepton of mass $m_{l_i}$ and a neutrino which for our purposes can
be approximated as massless.
The free energy per unit volume for the system in equilibrium
at temperature $T$ is given by
\begin{equation}
\label{Eq3-2}
\mathcal{F} = 6 \sum_{i=1}^{6} F \left( m_{q_{i}}, \mu \right)
+ \sum_{i=1}^{3} [2F \left( m_{l_{i}}, \mu_{i} \right) + F(0, \mu_{i})] \, ,
\end{equation}
where the $SU(2)_L$ gauge interactions maintain equilibrium between the charged
lepton and its neutrino and the up- and down-type quarks of a given
generation.
The free energy density for a (single helicity of a)
fermion of mass $m$ and chemical potential $\mu$ is given by
\begin{equation}
\label{Eq3-3}
F(m, \mu) = - T \int \frac{d^{3}K}{(2 \pi)^3}
[\ln(1+ e^{-(E+\mu)/T}) + \ln(1+ e^{-({E} - \mu)/T})] \, ,
\end{equation}
where $K_i$ is the spatial momentum of the fermion, and
$E \equiv \sqrt{K_i^2 + m^2}$ is its energy.
At high temperatures, $T \gg m, \mu$, this may be approximated as,
\begin{equation}
\label{Eq3-6}
F \left( m, \mu \right) \approx F \left( m, 0 \right)
- \frac{1}{12} \mu^{2} T^{2}
\left( 1 - \frac{3}{2 \pi^2} \frac{m^2}{T^2} \right) \, .
\end{equation}
The (individual) leptonic and baryonic number densities may written
\bea
\label{Eq3-4}
L_{i} & = & \frac{d}{d \mu_{i}} [2F(m_{i}, \mu_{i}) + F(0, \mu_{i})]
\approx - \frac{1}{2} \mu_i T^2 \beta_i \, , \\
\label{Eq3-5}
B & = & 2 \frac{d}{d \mu} \sum_{i=1}^{6} [F(m_{q_{i}}, \mu)]
\approx -\frac{1}{3} \mu T^2 \alpha \, ,
\eea
where
\begin{equation}\label{Eq3-8}
\alpha \equiv 6 - \frac{3}{2 \pi^2} \sum_{i=1}^{6}
\frac{m_{q_{i}}^2}{T^2}, ~~~~~ \beta_{i} \equiv 1 -
\frac{1}{\pi^2} \frac{m_{l_{i}}^2}{T^2} \, .
\end{equation}

Electroweak Sphaleron transitions violate $\sum L_i$ and $B$, but preserve the
three combinations $\Delta_i \equiv L_i - B / 3$.  In terms of the chemical
potentials these are
\begin{equation}
\label{Eq3-7}
\Delta_{i} \equiv L_{i} - \frac{1}{3} B
\approx \frac{\mu T^2}{9} \alpha - \frac{\mu_{i} T^2 }{2} \beta_{i}  \, ,
\end{equation}
We can invert the above relations
to obtain each $\mu_i$ in terms of the
quark chemical potential $\mu$, temperature
$T$, and the conserved value of the corresponding $\Delta_i$.
Effectively, the EW sphalerons convert nine quarks and one lepton of
each family into nothing. In thermal equilibrium, this leads to the
relation $\mu = - \sum_{i} \mu_{i} / 9$.  Using this fact,
together with the three conservation equations Eq~(\ref{Eq3-7}), allows us
to express the quark chemical potential in terms of the values of
the $\Delta_i$,
\begin{equation}
\label{Eq3-9}
\mu = \Big( \frac{2}{T^2}\sum_{i=1}^{3} \frac{\Delta_{i}}{\beta_{i}} \Big)
\Big( 9 + \frac{2}{9} \sum_{i=1}^{3} \frac{\alpha}{\beta_{i}} \Big)^{-1}  \, ,
\end{equation}
which can be combined with Eq~(\ref{Eq3-4}) to obtain the
final baryon number density \cite{Kuzmin:1987wn}
\begin{equation}
\label{Eq3-13}
B = \left\{
\begin{array}{ll}
\displaystyle
\frac{4}{13} (B -L) & B - L \neq 0\,  \\
- \frac{4}{13 \pi^2}\sum_{i=1}^N
\Delta_i \frac{m_{l_i}^2}{T^2} & B - L =0 \ .
\end{array}\right.
\end{equation}
The first of these results is
the familiar relationship applicable to theories that directly generate
a non-zero $B-L$ (such as leptogenesis) and indicates that in such theories
primordial $B$ cannot be completely washed out,
and a primordial $L$ will be converted into $B$ by EW sphalerons.
The second result shows
how in a theory with $B=L=0$ but the individual $\Delta_i$ non-zero,
the turn on of the charged lepton masses will also generate a non-zero
$B$.  In the scenario we are considering, with initially
$B=0$ and $L_3 = 2 \Delta /3$,
and taking the freeze-out temperature to be the close to the EW scale,
the resulting baryon number
is diluted to about $B \sim 10^{-6} \Delta$~\cite{Kuzmin:1985mm,
Kuzmin:1987wn,Dreiner:1992vm}.

Since the dilution factor plays a relevant role in our work, let us expand
on its origin: To compute the above quoted dilution factor, we have
assumed a second order phase electroweak phase transition.
Under this condition, the sphaleron processes will remain in
equilibrium until the weak spahleron rate is of the
order of the expansion rate of the Universe. The departure from
equilibrium therefore occurs at
the freeze-out temperature $T_F$, such that $v(T_F)/T_F \simeq 1$.
Using the relation $m_{\tau}(T) \simeq h_{\tau}/\sqrt{2} v(T)$,
and the condition $v(T_F)/T_F = 1$, we get
that the final baryon number is approximately given by
\begin{equation}
\label{eq:dilution} B \simeq -\frac{4}{13 \pi^2} \Delta
\frac{h_{\tau}^2}{2} \simeq - 1.6 \times 10^{-6} \Delta .
\end{equation}

\section{Lepton Number Generation in the Topflavor Model}
\label{tf-model}

\subsection{The Topflavor Model}

The gauge extension of the SM that we consider is based on the gauge group
$SU(3)_c\times SU(2)_1\times SU(2)_2\times U(1)_Y$.
While the $SU(3)_c$ and $U(1)_Y$ subgroups
remain the same as those of the SM, the $SU(2)_L$ group of SM is
expanded to a larger $SU(2)_1\times SU(2)_2$ in a flavor dependent
way. The fermion content of the model is identical to SM. Under
the new $SU(2)_1$ and $SU(2)_2$ groups, the doublets of the third
generation transform as doublets under $SU(2)_1$ and singlets
under $SU(2)_2$, while the first and second generation doublets
are singlets of $SU(2)_1$ and doublets of $SU(2)_2$.  Thus,
their $SU(2)_1\times SU(2)_2 \times U(1)_Y$ quantum numbers are
\bea
\label{2-1}
Q^3 = (\mathbf{2}, \mathbf{1})_{1/6}, & &
Q^{1,2} = (\mathbf{1}, \mathbf{2})_{1/6}, \nonumber\\
L^3 = (\mathbf{2}, \mathbf{1})_{-1/2}, & &
L^{1,2} = (\mathbf{1}, \mathbf{2})_{-1/2}.
\eea

After symmetry breaking, the Standard Model $SU(2)_L$ group
emerges as the unbroken diagonal subgroup of $SU(2)_1\times SU(2)_2$.
The corresponding $SU(2)_L$ gauge coupling is
\begin{equation}\label{2-3}
g_L = \frac{g_1g_2}{\sqrt{g_1^2+g_2^2}}.
\end{equation}
This relation implies that when one of the gauge couplings becomes
large, the other one approaches $g_L$ from above, and thus both
$g_1$ and $g_2$ are necessarily larger than $g_L$.  Thus, a convenient
parameterization is given by,
\bea
g_1 \equiv \frac{g_L}{\sin \phi}, & &
g_2 \equiv \frac{g_L}{\cos \phi},
\eea
in terms of an angle $\phi$. We will use a simplified notation
$s \equiv \sin \phi$ and $c \equiv \cos \phi$ below.

The symmetry breaking of the extended gauge group, $SU(2)_1 \times SU(2)_2$
to the SM weak gauge group $SU(2)_L$ is accomplished by
introducing a vacuum expectation value~(VEV) to a scalar
field $\Sigma$, which transforms as a bidoublet $(\mathbf{2},\mathbf{2})$ under
the extended gauge group transformations.
After the
$SU(2)_1 \times SU(2)_2$ breaking, $\Sigma$
can be decomposed under the residual diagonal $SU(2)_L$ symmetry into
a complex singlet $\sigma$ and a complex triplet $\tau$
(half of which is eaten by the $Z^\prime$ and $W^\prime$s),
\bea
\label{3-2}
\Sigma = \frac{1}{2} \left(
\begin{array}{cc}
\sigma + \tau_3 & \sqrt{2} \tau_+ \\
\sqrt{2} \tau_- & \sigma - \tau_3
\end{array} \right) ~.
\eea
We introduce the scalar potential,
\begin{eqnarray}
\label{2-4}
V_{\Sigma} &=& m^2 |\Sigma|^2 + \lambda |(\Sigma \Sigma)|^2
+ \lambda^\prime |\Sigma|^4 + (
-\frac{1}{2}D (\Sigma \Sigma)
+ \widetilde{\lambda} (\Sigma \Sigma) |\Sigma|^2 + h.c. ~),
\end{eqnarray}
where $D$ and $\widetilde{\lambda}$
are complex parameters and we use a notation
that supresses the gauge indices:
$( \Sigma \Sigma) \equiv \Sigma_{a \bar{b}} \Sigma_{c \bar{d}}
\epsilon_{ac}\epsilon_{\bar{b}\bar{d}}$
and $|\Sigma|^2 \equiv \rm{Tr}(\Sigma^{\dag}\Sigma)$.
For appropriate choices of parameters, this potential results in the VEV,
\bea
\label{2-2}
\langle \Sigma_{i\bar{k}} \rangle & = &
\frac{1}{2} u_0 \; e^{i\theta_0} \delta_{i\bar{k}}.
\eea
which will generally be complex, and will be solution of the equations,
\bea
u_0^2 &=& \frac{D e^{2i \theta_0} + D^* e^{-2i \theta_0} - m^2}
{\lambda + \lambda^\prime + \widetilde{\lambda} e^{2i \theta_0}
+ \widetilde{\lambda}^* e^{-2i \theta_0} }\\
\theta_0 & = & - \frac{1}{4} {\rm acos} \:{\rm Re}\: \left[
\frac{-2D^* + \widetilde{\lambda}^* u_0^2} {-2D +
\widetilde{\lambda} u_0^2} \right] ~. \eea Choosing some
representative parameters, taking $m = 200$ GeV, $D = 5 \times
10^5~e^i$ GeV$^2$, and
$\lambda=\lambda^\prime=\widetilde{\lambda}= 0.05$, we obtain a
zero temperature VEV described by $u_0 \simeq 2.7$ TeV and
$\theta_0 \simeq -0.7$. 
This particular set of parameters has been chosen with small quartic
interactions in order to have a first order phase transition, with the
dimensionful quantities arranged such that the $SU(2) \times SU(2)$
breaking scale is of order TeV.
Precision electroweak constraints have
been extensively considered in the literature
\cite{Chivukula:1995gu}, and typically require $u_0 \geq$ a few
TeV. Requiring that the extended instantons of the strongly
coupled $SU(2)$ do not mediate unacceptably large proton decay
\cite{Morrissey:2005uz} further requires the gauge couplings to
satisfy $\sin^2 \phi \gtrsim 0.2$. We will illustrate our results
with the representative point chosen above, and $s^2 =0.4$, in
order to be consistent with all constraints.

The fermion doublets of either $SU(2)_1$ or $SU(2)_2$ transform as doublets
under $SU(2)_L$. In addition, there is a $SU(2)_L$ triplet
of heavy gauge bosons from the breaking.
We denote the neutral and charged heavy gauge bosons as $Z^\prime$
and ${W^\prime}^{\pm}$.  Their masses are degenerate and given by
\begin{equation}
\label{2-3a}
M_{{W^\prime}^{\pm}} = M_{Z^\prime} = (g_1^2 + g_2^2) u^2
= \frac{g_L^2}{s^2 c^2} u^2.
\end{equation}

At the electroweak scale, $v\simeq 174$~GeV, the remaining $SU(2)_L\times
U(1)_Y$ electroweak symmetry is broken to $U(1)_{em}$ as in the
SM.  This is accomplished by giving a VEV to one or more Higgs
boson doublets. There are two possible representations for the
Higgs boson under $SU(2)_1$ or $SU(2)_2$, either
$(\mathbf{2},\mathbf{1})_{\pm 1/2}$ or $(\mathbf{1},\mathbf{2})_{\pm 1/2}$.
We will focus on the first case (sometimes called the {\em heavy} case)
as it motivates the large third family fermion masses
as they are the only family whose Yukawa interactions are
$SU(2)_1 \times SU(2)_2$ gauge-invariant.
In a non-supersymmetric theory, a single $Y=+1/2$ Higgs doublet
$H_{u}$ suffices,
but our results are largely unchanged if we generalize to a $Y=\pm 1/2$
pair of doublets, $H_u$, $H_{d}$ instead.  In order to connect more easily
with the supersymmetric case, we consider the case with two Higgs doublets
below.

The Yukawa couplings for the first two generations
can be generated by adding an additional ``spectator''
Higgs-like doublet $H^\prime_{u}$
(in a supersymmetric theory it would be a pair of doublets including
$H^\prime_{d}$) charged under $SU(2)_{2}$.
They couple to the first two
generations via Yukawa couplings and mix (slightly) with the regular Higgs(es)
via interactions such as $A_{1} H_{u} \Sigma H^\prime_{d}$. The small
Yukawa couplings for the first two generations can be naturally
obtained from the small mixing angle between $H^\prime$ and $H$.
The full set of this kind of gauge-invariant $\Sigma$-$H$-$H^\prime$
interactions between Higgs(es), spectator Higgs(es), and $\Sigma$ include,
\bea
\label{2-6}
A_{1} H_{u} \Sigma H^\prime_{d} + h.c; ~~~~~~
A_{2} H_{d} \Sigma H^\prime_{u} + h.c, \nonumber \\
A^\prime_{1} H_{u} \Sigma^{\dag} H^\prime_{d} + h.c; ~~~~~~
A^\prime_{2} H_{d} \Sigma^{\dag} H^\prime_{u} + h.c, \nonumber \\
c_{2} \mu^{*} H^\prime_{u} \Sigma H^{\dag}_{u} + h.c; ~~~~~~
c_{1}^{*} \mu^\prime H^\prime_{u} \Sigma^{\dag} H^{\dag}_{u} + h.c,
\nonumber \\
c_{1} \mu^{*} H^\prime_{d} \Sigma H^{\dag}_{d} + h.c; ~~~~~~
c_{2}^{*} \mu^\prime H^\prime_{d} \Sigma^{\dag} H^{\dag}_{d} + h.c
\eea
These interactions will be important
below, because they (indirectly) drive the Topflavor
phase transition's production of $L_3$.

\subsection{The Phase Transition}
\label{phase transition}

The details of the $SU(2)_1 \times SU(2)_2 \rightarrow SU(2)_L$
phase transition depend on the finite temperature effective
potential.  In order to generate a baryon asymmetry in the
Topflavor model, the processes which violate baryon- and
lepton-number must be out of equilibrium at the time of the phase
transition.  We are interested in the regime where $g_1 \gg g_2$,
for which we can approximate the sphalerons associated with the
$SU(2)_1 \times SU(2)_2$ symmetry breaking as purely arising from
$SU(2)_1$.   Their rate, \bea \Gamma_{sph} & \simeq & \kappa \:
\frac{u^7}{T^6}  \: {\rm Exp} \left[ - E_{sph} / T \right] ~, \eea
(with $E_{sph} = 4 \pi u / g_1$ and $\kappa$ a dimensionless
parameter of order one \cite{Huet:1995sh}) must be much less than
the Hubble expansion, $H \sim g_*^{1/2} T^2 / m_{Pl}$.  Assuming
$\kappa \sim 1$, $g_* \sim 100$ and $u_c \gtrsim T_c \sim~{\rm
TeV}$, this requires, \bea \frac{u_c}{g_1 T_c} & \gtrsim & 2.5 ~,
\eea and for $s^2 \sim 0.4$, we should have\footnote{Note that
such a strong first order phase transition may provide a strong
signature in gravitational waves \cite{Kosowsky:1992rz,Apreda:2001us, Nicolis:2003tg, Grojean:2006bp, Randall:2006py}
detectable at the planned space interferometer, LISA. We will
pursue this idea in a separate paper\cite{STW}. } $u_c / T_c
\gtrsim 2.5$.

The potential for $\Sigma$, Eq.~(\ref{2-4}), can be expanded to,
\bea
\label{3-1}
V_0(\sigma) = \frac{1}{2} m^2 |\sigma|^2
+ \frac{\lambda + \lambda^\prime}{4} |\sigma|^4
+ \left( -\frac{1}{2} D \sigma^2
  + \frac{\widetilde{\lambda}}{4} \sigma^2 |\sigma^2| + h.c.\right)
\eea where for brevity we have not shown the terms involving the
triplet component $\tau$. At one loop, there are both
temperature-dependent and temperature-independent corrections to
the potential, \bea \label{3-6} V(u, \theta, T) &=& V_0(u, \theta,
0) + V_1(u, \theta, 0) + V_1(u, \theta, T) ~. \eea We consider the
limit where the gauge couplings of the $SU(2)$'s are much larger
than that of the $\Sigma$ self-interactions $\lambda$ and thus
approximate the complete one-loop corrections by those from the
gauge bosons, $Z^\prime$ and $W^\prime$. We renormalize parameters
according to a scheme that absorbs the correction to the position
of the zero-temperature minimum, and find, \bea \label{3-4} V_1(u,
\theta, 0) &=& \frac{9}{64 \pi^2} \left( \frac{g^2_L}{s^2 c^2}
\right)^2 u^2 \left[ u^2 \left(\log \frac{u^2}{u_0^2} -
\frac{3}{2} \right) + 2  u_0^2 \right] \eea and finite temperatue
correction, \bea \label{3-5} V_1(u, \theta, T) &=& \frac{g_i
T^4}{2 \pi^2} I \Big( \frac{m_i(\sigma)}{T} \Big) \eea with \bea I
\Big( \frac{m_i(\sigma)}{T} \Big) &=& \int_0^\infty dx \cdot x^2
\left\{ \log \left( 1- e^{-\sqrt{x^2 + g^2 u^2/ (s^2 c^2
T^2)}}\right) - \log(1-e^{-x}) \right\}, \eea where $g_i = 9$. The
Debye screening effect on the longitudinal modes of massive gauge
bosons is neglected because their corrections $g_1 T$ is small
compared to their masses induced by the Higgs
vev\cite{Dine:1992wr}. Note that the one loop corrections from the
gauge sector depend only on the magnitude of the VEV $u$ and not
on its phase.

\begin{figure}[t]
\begin{center}
\psfig{figure=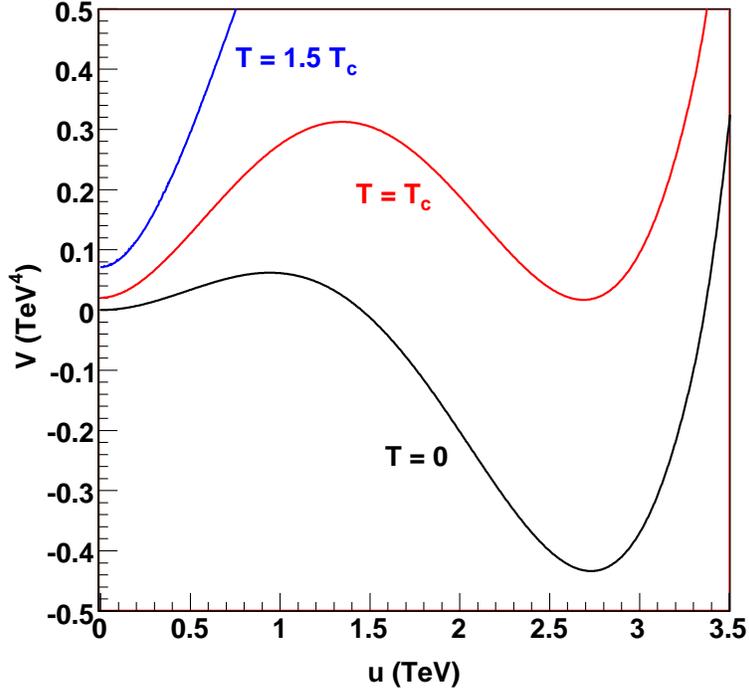,height=10cm}
\caption{The effective potential as a function of the magnitude of the
VEV for three different choices of temperature.  The phase of the VEV at
each point is chosen as the solution of the Equations of motion for that
value of the magnitude.}
\label{potential-Tc}
\end{center}
\end{figure}

For the sample parameters chosen above,
$m = 200$ GeV, $D = 5 \times 10^5~e^i$ GeV$^2$,
$\lambda=\lambda^\prime=\widetilde{\lambda}=0.05$,and $s^2 = 0.4$,
we find that the critical temperature for these parameters is
$T_c \simeq 840$~GeV, and the VEV at $T_c$ is described by
$u_c \simeq 2.7$~TeV, $\theta_c \simeq -0.7$,
indicating a first order phase transition that is easily strong enough.
In Figure~\ref{potential-Tc}
we plot the effective potential for several choices of temperature.

\subsubsection{Bubble Profile and Evolution}

At $T_c$, the rate for nucleation of
bubbles with $\langle \sigma \rangle \neq 0$ becomes large, and the bubbles
expand to fill the universe with the true vaccuum.  In this subsection
we make some rough estimates of the properties of the nucleated bubbles,
which are pertinent to the eventual generation of baryon asymmetry as
they provide the out-of-equilibrium dynamics which results in lepton-number
being unequally distributed through-out the three generations.
We will simplify the treatment by considering the phase transition as a
quasi-equilibrium process such that the temperature is varying slowly enough
that the properties of the nucleating bubbles can be obtained by studying
fixed temperature solutions at $T \simeq T_c$.

\begin{figure}[htbp]
  \begin{center}
    \psfig{figure=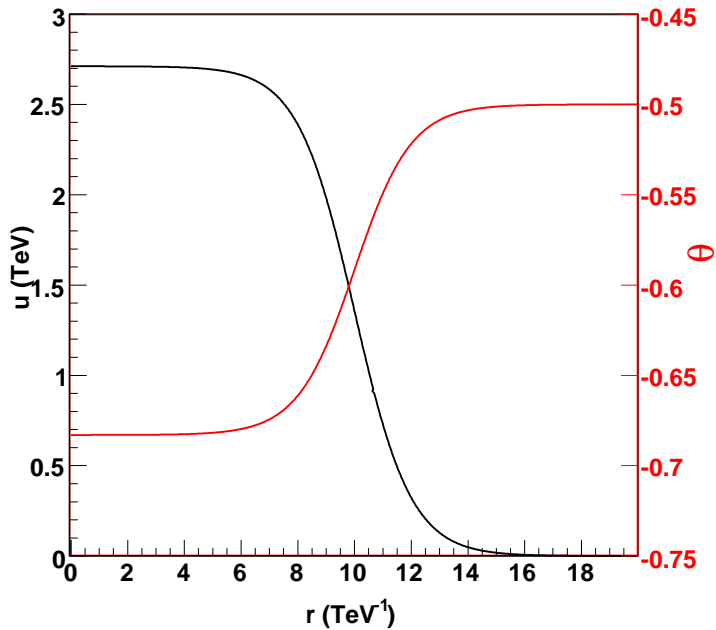,height=9cm}
  \caption{The bubble profile at the critical temperature $T_c$ for
           $R = 10~{\rm TeV}^{-1}$.}
  \label{Wall-width}
  \end{center}
\end{figure}

Assuming that the nucleated bubbles have spherical symmetry
\cite{Col78}, the Euclidean action of the configuration
becomes\footnote{We use the $O(3)$ approximation since the
nucleation temperature $T_n \simeq T_c> 1 /
(2R)\cite{Linde:1981zj}$.} \bea \label{accion} S_3(T) &=& 4 \pi
\int dr\; r^2 \Big\{ \Big( \nabla \langle \sigma \rangle \Big)^2 +
V(\langle \sigma \rangle, T_c) \Big\} \eea where $\langle \sigma
\rangle$ is a function of radius $r$ that describes the
configuration.  The transitions will be predominantly mediated by
the solutions which minimize this action, give by the solutions of
the equations, \bea
\frac{d^2 u} {d r^2} + \frac{2}{r} \frac{d u}{d r}
& = & \frac{\delta V}{\delta u} \nonumber\\
u^2 \left[ \frac{d^2 \theta} {d r^2} + \frac{2}{r} \frac{d \theta}{d r}
\right] & = & \frac{\delta V}{\delta \theta}
\label{burbuja}
\eea
subject to the boundary conditions,
\bea
\left.\frac{du}{d r} \right|_{r=0} & = & 0;
\qquad \left. u \right|_{r \rightarrow \infty} = 0
\nonumber \\
\left.\frac{d\theta}{d r} \right|_{r=0} & = & 0;
\qquad \left. \theta \right|_{r \rightarrow \infty} = \theta_{u=0}
\: .
\label{frontera}
\eea
These coupled differential equations are somewhat difficult to solve.  Instead
of looking for detailed solutions, we will use an ansatz for the profile, and
a variational approach to determine the parameters that describe it.  We
write the solutions in the form of two ``kinks'' with the proper
asymmptotic behavior,
\bea
u (r) & = & \frac{u_c}{2} \left[
1 - {\rm Tanh} \left( \alpha \left( r - R \right) \right) \right]
\nonumber \\
\theta (r) & = & \theta_{u=0} + \frac{\theta_c - \theta_{u=0}}{2}
\left[ 1 - {\rm Tanh} \left( \alpha \left( r - R \right) \right)
\right] \label{eq:ansatz} \eea where the radius of the bubble is
$\sim R$ and the width of the bubble wall is $\sim 1 / \alpha$. In
addition to imposing the form of the solution, we also assume that
the bubble width is the same for $u$ and for $\theta$ (which we
expect to be approximately true). We determine $\alpha$
numerically by plugging the solutions of Eq.~(\ref{eq:ansatz})
into the action Eq.~(\ref{accion}) and finding the value of
$\alpha$ which minimizes the action.  For our standard parameter
choice we find $\alpha \sim T / 2$ and $L_w \equiv 1/ \alpha \sim
2 / T$. The profiles are plotted in Fig~\ref{Wall-width}. We
expect that for large bubbles, the details will become independent
of $R$, which in fact proves to be true.

Under our quasi-equilibrium assumption, the expansion of the
bubble is driven by the fact that the gauge bosons acquire masses
inside the bubble, and thus the free energy is minimized for large
bubbles \cite{Dine:1992wr,Moore:1995si}. By analogy with the SM EW phase
transition, we estimate the bubble wall velocity $v_w \sim 0.05$,
though we find that our final results are very insensitive to the
precise value of $v_w$.

\subsection{Diffusion Equations in the Topflavor Model}
\label{diffusion equation}

We now compute the prediction for the $L_3$ generated during the transition
in which the Topflavor model breaks down to the SM.  The underlying picture
is similar to  the standard EWBG picture in the SM (or MSSM).  The
bubble of true vacuum is expanding, and generates chiral charge through
the $CP$-violating interaction of the plasma with the bubble wall.
In the specific case of Topflavor, the particles which interact
strongly with the wall are the Higgses, through the interactions
in Eq~(\ref{2-6}).
These charges diffuse freely in the unbroken phase and are converted into
$B$ and $L_3$ by a combination of the Yukawa interactions and
the unsuppressed sphalerons.  As they pass into the
broken phase, they are frozen.

In the limit $g_1 \gg g_2$, we neglect the $SU(2)_2$ sphalerons
associated with the first two families.  The quark Yukawa
interactions and the QCD instantons, together with the fact that
all of the quarks diffuse at approximately the same rate, allows
us to constrain the light quark densities in terms of the
right-handed bottom density $b$, \bea \label{5-1} Q_{1L} = Q_{2L}
= -2 U_R=-2 D_R=-2 S_R=-2 C_R = -2 b \, . \eea Thus, the species
whose densities we will track are the left-handed top and bottom
doublet, $Q \equiv t_L + b_L$, the right-handed top $t \equiv
t_R$, the right-handed bottom $b\equiv b_R$, the left-handed
lepton doublet $L_3 \equiv \tau_L + \nu_{\tau}$, and the Higgs $h
\equiv (h^+_u + h^0_u - h^-_d - h^0_d )$. We assume that the
$H$-$H^\prime$-$\Sigma$ interactions Eq.~(\ref{2-6}) are fast
enough such that the spectator Higgses $H^\prime$ are kept in
equilibrium with the Higgs, and thus $\Sigma=0$, $h^\prime \equiv
({h^\prime}_u^+ + {h^\prime}_u^0 - {h^\prime}_d^- -
{h^\prime}_d^0) = h$, and we do not include the densities $\Sigma$
and $h^\prime$ in the diffusion equations. For relativistic
particles near equilibrium, we can write the number densities in
terms of a chemical potential $\mu_i$ as $n_i = k_i \mu_i T^2 / 6$
where $k_i$ counts the number of degrees of freedom, \bea
\label{5-c} k_{Q} = 6;~~~ k_{L} = 2; ~~~k_{t}=k_{b}=3; ~~~k_{h}=8
\, . \eea

The diffusion equations will contain the interactions which are fast
compared to the time scales at which the elements of the plasma are
diffusing, $\tau_i = D_i / v_w$, where $v_w \sim 0.05$
is the speed of the bubble wall's expansion and $D_i$ is a diffusion
coefficient which characterizes the interactions with the background
plasma.  Typically, one expects $D_Q = D_t = D_b \simeq 6 / T$ and
$D_L \simeq D_h \simeq 110 / T$ \cite{Joyce:1994zn}.
Thus, we consider the
processes characterized by rate $\Gamma \gtrsim \tau_Q$.  These interactions
include the $SU(2)_1$ sphalerons with rate $\Gamma_1$,
the QCD instantons with rate
$\Gamma_{QCD}$, and the top quark Yukawa coupling to
the Higgs with rate $\Gamma_y$.
We continue to assume that the sphalerons associated
with $SU(2)_2$ can be neglected.  These rates
are estimated to be equal to~\cite{Huet:1995sh,Arnold:1996dy},
\bea
\Gamma_y & \simeq &
\frac{27}{2}\lambda_t^2 \alpha_S
\left( \frac{\zeta (3)}{\pi^2} \right)^2 T = 7.4~{\rm GeV} \, , \\
\Gamma_{QCD} & \simeq &
16 \kappa^\prime \alpha_S^4 T = 0.3~{\rm GeV}  \, , \\
\Gamma_1 & \simeq &  30 {\alpha_1}^5 T = 0.1~{\rm GeV} \, ,
\eea
where $\lambda_t \sim 1$ is the top Yukawa coupling, $\alpha_S(T_c) \sim 0.08$
is the strong coupling constant, and $\kappa^\prime \sim 1$ is a
dimensionless coefficient.
We have evaluated the rates for
$s^2 = 0.4$ and $T_c = 840$ GeV, as is appropriate for our example parameter
set.

We approximate the bubble as large, and thus treat the problem
one-dimensionally, with the $z$-axis perpendicular to the wall,
whose location is at $z=0$, with the $z >0$ side in the broken
phase. The rates of change of the various densities are described
by the coupled set of equations\footnote{Note that leptons
diffuse faster than quarks, and thus the $B-L$ charge density
$(Q+t+b)/3 - L$ is not zero locally.}
 \bea v_w Q^\prime - D_Q Q^{\prime \prime}
&=& -\Gamma_y \left[ \frac{Q}{k_{Q}} - \frac{h}{k_{h}} -
\frac{t}{k_{t}} \right] - 6 \Gamma_{QCD} \left[ 2\frac{Q}{k_{Q}} -
\frac{t}{k_{t}} - 9 \frac{b}{k_{b}} \right] \nonumber \\ & & - 6
\Gamma_{1} \left[ 3 \frac{Q}{k_{Q}} + \frac{L_3}{k_{L}} \right]
 \, , \nonumber \\
v_w t^\prime - D_Q t^{\prime \prime}  &=&
-\Gamma_y \left[- \frac{Q}{k_{Q}} + \frac{h}{k_{h}} + \frac{t}{k_{t}} \right]
+ 3 \Gamma_{QCD} \left[ 2\frac{Q}{k_{Q}}
- \frac{t}{k_{t}} -9\frac{b}{k_{b}} \right]
 \, , \nonumber \\
v_w h^\prime - D_h h^{\prime \prime} & = &
-\Gamma_y \left[- \frac{Q}{k_{Q}} + \frac{h}{k_{h}} + \frac{t}{k_{t}} \right]
+ \gamma_{h}
 \, ,  \nonumber \\
v_w b^\prime - D_Q b^{\prime \prime} & = &
3 \Gamma_{QCD} \left[ 2\frac{Q}{k_{Q}}
- \frac{t}{k_{t}} -9\frac{b}{k_{b}} \right]
 \, , \nonumber \\
v_w L_3^\prime - D_L L_3^{\prime \prime} & = &
-2 \Gamma_{1} \left[ 3 \frac{Q}{k_{Q}} + \frac{L_3}{k_{L}} \right] \, ,
\label{5-2}
\eea
where primes denote derivatives with respect to $z$ and
$\gamma_h$ is the $CP$-violating source for the Higgs induced by
the bubble wall, approximated as a step function,
\begin{equation}\label{5-d}
\gamma_{h} = \left\{
  \begin{array}{lr}
  \displaystyle \tilde{\gamma}_{h_u} - \tilde{\gamma}_{h_d}~~~~~~~
  & (-L_w<z \leq 0) \\
  \displaystyle 0 &  (z>0 ~~ \textrm{or} ~~ z<-L_w)
  \end{array}\right. \: ,
\end{equation}
whose magnitude we estimate below.

\subsection{$CP$-violating Sources from Spontaneous $CP$ violation}
\label{CPV}

We consider the $CP$-violation arising from the spontaneous
$CP$-violation associated with the phase of the VEV $\langle
\sigma \rangle$.  This field couples directly to the EWSB Higgses
$H_u$ and $H_d$ and thus influences their number densities as they
scatter off of the bubble wall.  As we saw in section~\ref{phase
transition}, the phase of $\langle \sigma \rangle$ varies as one
moves from inside the bubble of true vacuum to the unbroken phase
(see i.e. Figure~\ref{Wall-width}), \bea \label{4-2a} \Delta
\theta & \equiv & \theta_c - \theta_{u=0} = - \frac{1}{4} {\rm
acos} \left[ \frac{-2D^* + \widetilde{\lambda}^* u_c^2} {-2D +
\widetilde{\lambda} u_c^2} \right] +\frac{1}{2} \eea and thus is
space-time dependent as the bubble expands.

In computing the value of the $CP$-violating source $\gamma_h$, we follow
the treatment first introduced by Riotto~\cite{Riotto:1995hh,Riotto:1997vy}
based on the
closed time path (CTP) formalism, which allows us to capture the main
non-equilibrium quantum effects.
The CTP formalism distinguishes fields
with arguments on the positive and negative
branches of the closed time path. This doubling of fields
leads to six different real-time propagators; for a generic scalar
field $\phi$,
\begin{eqnarray} \label{4-3a}
G_{\phi}^{>}(x,y) &=& -i \langle \phi(x) \phi^{\dag}(y) \rangle,
\nonumber \\ G_{\phi}^{<}(x,y) &=& -i \langle \phi^{\dag}(x)
\phi(y) \rangle, \nonumber \\ G_{\phi}^{t}(x,y) &=& \theta(x,y)
G_{\phi}^{>}(x,y) + \theta(y,x) G_{\phi}^{<}(x,y), \nonumber \\
G_{\phi}^{\bar{t}}(x,y) &=& \theta(x,y) G_{\phi}^{<}(x,y) + \theta(y,x)
G_{\phi}^{>}(x,y) ,
\end{eqnarray}
which are conveniently written as a matrix $\title{G(x,y)}$:
\be
\label{4-3b}
\widetilde G(x,y)= \left(\begin{array}{cc}
G^t(x,y) & -G^<(x,y) \\
G^>(x,y) & -G^{\bar t}(x,y)
\end{array}\right).
\ee
{}From the Schwinger-Dyson equations of the path-ordered two-point
functions, one obtains
\begin{eqnarray}
\label{4-3}
\partial_{\mu}j^{\mu}_{\phi} &=&
- \int d^3 z \int^{X_{0}}_{- \infty} dz_{0}
\left[ \Sigma^{>}_{\phi}(X, z) G^{<}_{\phi}(z, X) - G^{>}_{\phi}(X, z)
\Sigma^{<}_{\phi}(z, X) \right.
\nonumber \\ & & \left.
+ G^{<}_{\phi}(X, z) \Sigma^{>}_{\phi}(z, X) - \Sigma^{<}_{\phi}(X,z)
\Sigma^{>}_{\phi}(z, X) \right] .
\end{eqnarray}

\begin{figure}[t]
\begin{center}
\psfig{figure=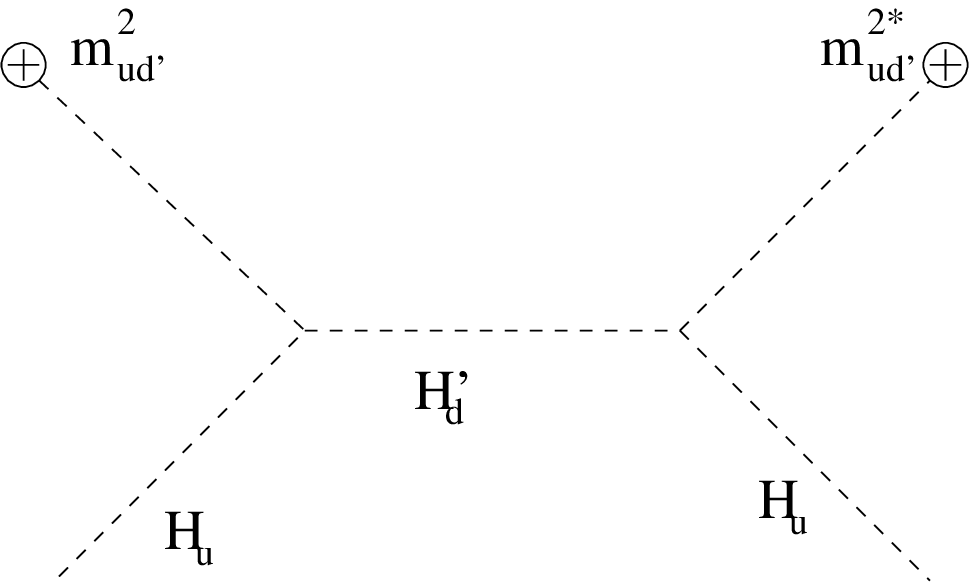,height=4cm} \hspace*{1cm}
\psfig{figure=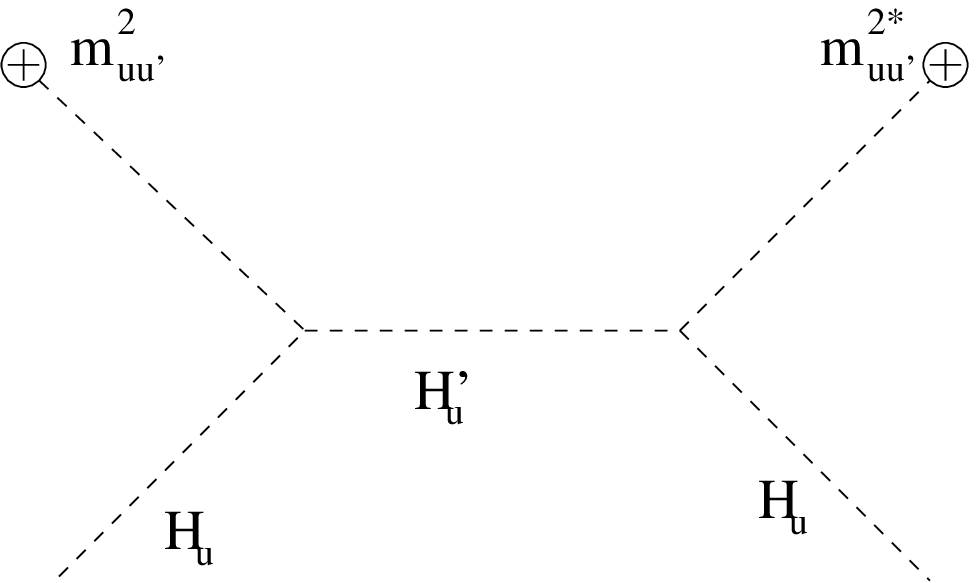,height=4cm} \caption{Feynman diagrams for the
leading contributions to the self-energies of $H_u$ in the
background of a spacetime varying $\langle \Sigma \rangle$.}
\label{fig:selfe}
\end{center}
\end{figure}

The leading contribution to the self energies
$\tilde{\Sigma}_{H_u}(x, y)$ and $\tilde{\Sigma}_{H_d}(x, y)$
from the interactions of Eq.~(\ref{2-6}) are (see Figure~\ref{fig:selfe})
\begin{eqnarray}
\tilde{\Sigma}_{H_u}(x, y) & = & g(x, y) \tilde{G}_{H^\prime_u}(x, y)
+ h(x,y) \tilde{G}_{H^\prime_d}(x, y),
\nonumber \\
\tilde{\Sigma}_{H_d}(x, y) & = & \tilde{g}(x, y) \tilde{G}_{H^\prime_d}(x, y)
+ \tilde{h}(x,y) \tilde{G}_{H^\prime_u}(x, y),
\label{4-5}
 \end{eqnarray}
where
\begin{eqnarray}
\label{4-6} g(x,y) & = & m^2_{u u'}(x) {m^2}^*_{u u'}(y)= \left[
c_2^* \mu \Sigma^{\dag}(x) + c_1 {\mu^\prime}^{*} \Sigma(x)
\right] \left[ c_2 \mu^* \Sigma(y) +
c_1^* {\mu^\prime} \Sigma^{\dag}(y) \right] \nonumber \\
h(x,y) & = & m^2_{u d'}(x) {m^2}^*_{u d'}(y)= \left[ A_1 \Sigma(x)
+ A^\prime_1 \Sigma^{\dag}(x) \right] \left [A_1^*
\Sigma^{\dag}(y) + {A^\prime}_1^* \Sigma(y) \right]
\nonumber \\
\tilde{g}(x,y) & = & m^2_{d d'}(x) {m^2}^*_{d d'}(y)= \left[ c_2
{\mu^\prime}^{*} \Sigma (x) + c_1^* {\mu} \Sigma^{\dag} (x)
\right] \left[ c_2^* {\mu^\prime} \Sigma^{\dag} (y) + c_1
{\mu}^{*} \Sigma (y) \right]
\nonumber \\
\tilde{h}(x,y) & = & m^2_{d u'}(x) {m^2}^*_{d u'}(y)=\left[ A_2
\Sigma(x) + A^\prime_2 \Sigma^{\dag}(x) \right] \left [A_2^*
\Sigma^{\dag}(y) + {A^\prime}_2^* \Sigma(y) \right]
 \end{eqnarray}
The $CP$-violating part of Eq.~(\ref{4-3}), evaluated for $H_u$,
is the imaginary part of Green functions,
\begin{eqnarray}
\label{4-7}
\gamma_{h_u} &=& - i \int d^3 z \int^{X_{0}}_{- \infty} d z_{0}
\Big \{ \left[ g(x, z)- g(z, x) \right]
\rm{Im} \left[ G_{H^\prime_u}^{>}(x, z) G_{H_u}^{<}(z, x)
- G_{H^\prime_u}^{<}(x, z) G_{H_u}^{>}(z, x)\right]
\nonumber \\ & &  +
\left[ h(x, z)- h(z, x) \right] \rm{Im}\left[ G_{H^\prime_d}^{>}(x, z)
G_{H_u}^{<}(z, x) - G_{H'_d}^{<}(x, z) G_{H_u}^{>}(z, x) \right] \Big \}.
 \end{eqnarray}
We can express the scalar Green's function in terms of the
Bose-Einstein distribution and the spectral density of a scalar field
\begin{eqnarray}
\label{4-7a}
G^ {\gtrless}(x, y) & = & \int \frac{d^4 k}{(2 \pi)^4} e^{-ik \cdot
(x-y)} g_B^\gtrless (k_0, \mu_i) \rho (k_0, \vec{k}),
\end{eqnarray}
where the equilibrium distribution functions are:
\bea
\label{4-7b}
g_B^{>}(\omega, \mu) & = & 1 + n_B(\omega, \mu_i) \\
\label{4-7c}
g_B^{<}(\omega, \mu) & = & n_B(\omega, \mu_i) ,
\eea
with $n_B(x) =1/[\exp(x/T) - 1]$. The spectral density in the limit of
small decay width is \cite{Riotto:1997vy},
\begin{eqnarray} \label{4-7d}
\rho (k_0, \vec{k}) &=& i \Big[ \frac{1}{(k^0 + i \varepsilon + i
\Gamma )^2 - \omega^2(|\vec{k}|)}
- \frac{1}{(k^0 - i \varepsilon - i \Gamma)^2 - \omega^2(|\vec{k}|)} \Big ],
\end{eqnarray}
where $\omega^2(|\vec{k}|) = \vec{k}^2 + M^2$. The mass $M$ and
width $\Gamma$ are the thermal mass and width, respectively.

The coefficients $[g(x, z)- g(z, x)]$ and $[h(x, z)- h(z, x)]$ can
be calculated to the first order in the expansion about $x=z$.
\begin{eqnarray} \label{4-8}
g(x,z) - g(z,x) &=& 2 i \left[ |c_1 \mu'|^2 - |c_2 \mu|^2 \right]
\sin \left[ \theta (x) -\theta (z) \right] u(x)u(z) \nonumber
\\ & \simeq &  2 i \left[ |c_1 \mu'|^2 - |c_2 \mu|^2 \right]
(x-z)^{\mu} \left[\partial_{\mu} \theta (x) \right] u^2(x) + \cdots.
 \end{eqnarray}
When these are inserted in Eq.~(\ref{4-7}), the fact that
the spectral density is isotropic in space, implies that only the
time component is non-vanishing, and we can make the replacements,
\begin{eqnarray}
\label{4-9}
g(x,z) - g(z,x) &=& 2 i \left[ |c_1 \mu|^2 - |c_2 \mu|^2 \right]
(x-z)^{0}] \left( \frac{\Delta \theta}{L_w}v_w \right) u^2(x), \\
\label{4-10}
h(x,z) - h(z,x) &=& 2 i \left[ |A_1|^2 - |A'_1|^2 \right]
(x-z)^{0} \left(\frac{\Delta \theta}{L_w}v_w \right) u^2(x) ,
\end{eqnarray}
and thus,
\begin{eqnarray}
\label{4-11}
\widetilde{\gamma}_{h_u} &=&
\left(\frac{\Delta \theta}{L_w}v_w \right) u^2(x)
\left\{ \left[ |c_1 \mu^\prime|^2 - |c_2 \mu|^2 \right]
\mathcal{I}_{H_u H_u^{^\prime}}
+ \left[ |A_1|^2 - |A^\prime_1|^2 \right]
\mathcal{I}_{H_u H_d^{^\prime}} \right\},
\end{eqnarray}
where
\begin{eqnarray}
\label{4-12}
\mathcal{I}_{H_u H^\prime_i} & = & \int_{0}^{\infty} dk
\frac{k^2}{2 \pi^2 \omega_{H^\prime_i} \omega_{H_u}} \times
\Big \{ (1+ 2 \rm{Re}[n_{B}(\omega_{H^\prime_i}+i \Gamma_{H^\prime_i})])
I(\omega_{H_u}, \Gamma_{H_u}, \omega_{H^\prime_i}, \Gamma_{H^\prime_i})
\nonumber \\ & &
+ (1+2\rm{Re}[n_{B}(\omega_{H_u}+ i \Gamma_{H_u})])
I(\omega_{H^\prime_i}, \Gamma_{H^\prime_i}, \omega_{H_u}, \Gamma_{H_u})
\nonumber \\ & &
- 2 (\rm{Im}[n_{B}(\omega_{H_u}+i \Gamma_{H_u})] +
\rm{Im}[n_{B}(\omega_{H^\prime_i}+i \Gamma_{H^\prime_i})]) G(\omega_{H_u},
\Gamma_{H_u}, \omega_{H^\prime_i}, \Gamma_{H^\prime_i}) \Big \}
 \, ,
\end{eqnarray}
and the functions $I$ and $G$ are given by
\begin{eqnarray} \label{4-13}
I(a, b, c, d) = \frac{1}{2}\frac{1}{[(a+c)^2 + (b+d)^2]}\sin
\Big [ 2 \arctan  \frac{a+c}{b+d} \Big] \nonumber \\
+ \frac{1}{2}\frac{1}{[(a-c)^2 + (b+d)^2]}\sin \Big [ 2 \arctan
\frac{a-c}{b+d} \Big] \nonumber \\
\end{eqnarray}
\begin{eqnarray} \label{4-14}
G(a, b, c, d) = - \frac{1}{2}\frac{1}{[(a+c)^2 + (b+d)^2]}\cos
\Big [ 2 \arctan \frac{a+c}{b+d} \Big] \nonumber \\ -
\frac{1}{2}\frac{1}{[(a-c)^2 + (b+d)^2]}\cos \Big [ 2 \arctan
\frac{a-c}{b+d} \Big]
 \, . \nonumber \\ & &
\end{eqnarray}

In exactly the same way, we derive,
\begin{eqnarray} \label{4-15}
\widetilde{\gamma}_{h_d} &=&
\left(\frac{\Delta \theta}{L_w}v_w \right) u^2(x)
\left\{ \left[ |c_1 \mu |^2 - |c_2 \mu^\prime |^2 \right]
\mathcal{I}_{H_d H_d^{^\prime}}
+ \left[ |A_2|^2 - |A^\prime_2|^2 \right]
\mathcal{I}_{H_d H_u^{^\prime}} \right\} .
\end{eqnarray}
The over-all magnitude of the $CP$-violating source depends
sensitively on several parameters which have up until now not
played a large role in deriving our ressults.  Thus, we content
ourselves with the order of magnitude estimate based on the sample
parameters for the $\sigma$ potential and bubble wall velocity and
profile, and assuming the thermal masses and widths are roughly
TeV, and that the $\Sigma$-$H$-$H^\prime$ dimensionful
interactions\footnote{The thermal widthes of the Higgs and
spectator Higgs are dominated by the trilinear interactions
$\Sigma$-$H$-$H^\prime$.} are of order TeV.  Evaluating the
integrals numerically, we find $\gamma_h \sim 0.01~{\rm TeV}~
\Delta \theta / L_w v_w u_c^2$, which for the choice of sample
parameters described above leads to $\gamma_h \sim 10^9$~GeV$^4$,
and acquires non-vanishing values only at the position of the
bubble wall, where the Higgs fields are varying.

The results presented above rely on a method similar to the one used
in Ref.~\cite{CQRVW} in the Minimal Supersymmetric Standard Model case.
Further improvements to this method were performed in Refs.~\cite{CMQSW1},
by considering the all-order resummation of the Higgs mass insertion
efffects. The result was a mild suppression of the results in the
case of degenerate masses, as well as new relevant terms away from
the degenerate mass regime. Furthermore, in Ref.~\cite{CQSW2}, 
based on self-consistency arguments, a more detailed
analysis of the relation between the CP-violating sources and the 
currents induced by the Higgs fields
was performed, leading to the presence of higher order derivatives in
the sources, as first suggested in Ref.~\cite{CJK}. The final result
of these investigations is a suppression of the baryon asymmetry
by a factor of a few in the
degenerate mass regime compared to the one obtained in Ref.~\cite{CQRVW}, 
as well as a power-law suppression with the
masses when they move away from the degenerate region.

One of the weaknesses of the above described work is the lack of
a rigorous derivation of the sources for the diffusion equation.
A derivation of the semiclassical forces in the transport equations
by means of the dynamics of the two-point function in the Schwinger-Keldysh
formalism was performed in 
Refs.~\cite{KPSW1,KPSW2,Prokopec:2003pj,Konstandin:2005cd}. 
In particular, in Ref.~\cite{Konstandin:2005cd} a consistent treatment of the
fermion mixing effects was performed, and a careful derivation of the sources
in the diffusion equations was obtained. The final result was 
smaller by a factor of a few to an order of magnitude
than the result obtained in Ref.~\cite{CQSW2} in the 
degenerate mass regime, and a stronger than power-law suppression 
away from the degenerate case.  In our work we are interested in an
order of magnitude estimate of the result for the baryon asymmetry, 
and for that purpose the results
of this section are sufficient.  However, while an enhancement of the sources
by a factor of a few would be easy to obtain by a careful choice of the
parameters of our model, an enhancement by several order of magnitudes
woulde prove very difficult. For these reasons, 
it would be interesting to pursue a more
general treatment using the techniques of Ref.~\cite{Konstandin:2005cd} 
to make a detailed
exploration of the full parameter space consistent with the observed
baryon asymmetry.

\subsection{Results}

We assemble the results and make a prediction for the density of lepton number
stored in the third family, $L_3$.
We have solved the diffusion equations in two ways; the first is a
brute force numerical solution of the coupled differential equations, while
the second proceeds by making some simplifying approximations which allow us to
determine an analytic solution to the diffusion equations.
Our analytic solution simplifies the problem by assuming that $\Gamma_y$,
$\Gamma_{QCD}$ and (for $z<0$) $\Gamma_1$ are all strong enough that they
enforce near-equilibrium relations among the number densities of
the species participating in the interactions.  Thus, we have,
\bea
\frac{Q}{k_Q} - \frac{h}{k_h} - \frac{t}{k_t}
\sim \frac{1}{\Gamma_y} \sim 0 \: , \nonumber \\
2\frac{Q}{k_Q} - \frac{t}{k_t} - 9 \frac{b}{k_b}
\sim \frac{1}{\Gamma_{QCD}} \sim 0 \: ,
\label{eq:const1}
\eea
and for $z<0$,
\bea
3 \frac{Q}{k_Q} + \frac{L_3}{k_L} \sim \frac{1}{\Gamma_1} \sim 0 .
\label{eq:const2}
\eea
Following the treatment of the usual EW case \cite{Huet:1995sh},
we take linear combinations of the diffusion equations (\ref{5-2}) which
are independent of $\Gamma_y$, $\Gamma_{QCD}$, and $\Gamma_1$ and use the
equilibrium relations to express the remaining two equations in terms of
$Q$ and $h$.  The result can be expressed as a matrix differential equation,
\bea
M \left[
\begin{array}{l} Q^{\prime \prime} \\ h^{\prime \prime} \end{array} \right]
+ N \left[ \begin{array}{l} Q^{\prime} \\ h^{\prime} \end{array} \right]
= \left[ \begin{array}{l} -\gamma_h \\ 0 \end{array} \right]
\label{eq:diffm}
\eea
where we simplfy $\gamma_h$ as constant for $-L_w < z < 0$ and $M$ and $N$ are
constant matrices, functions of the $D$'s, $k$'s, and $v_w$,
\bea
M & = &
\left[ \begin{array}{cc}
D_Q \left( \frac{k_b-9 k_t}{9 k_b}\right) &
D_h+ D_Q \left(\frac{9 k_t + k_b}{9k_h} \right) \\
-D_Q \left(\frac{9 k_Q + 9 k_t + k_b}{9 k_Q}\right)
- D_L \left( \frac{9 k_L}{k_Q} \right) &
D_Q \left( \frac{9 k_t - k_b}{9 k_h} \right)
\end{array} \right] =
\frac{1}{T} \left[ \begin{array}{cc}
-\frac{8}{3} & \frac{225}{2} \\
-\frac{28}{3} & 2
\end{array} \right] \: , \nonumber \\
N & = & v_w \left[ \begin{array}{cc}
\frac{9 k_t - k_b}{9 k_Q} & -\frac{9 k_h + 9 k_t + k_b}{9 k_h} \\
\frac{9 k_Q + 81 k_L + 9 k_t + k_b}{9 k_Q} & \frac{k_b - 9 k_t}{9 k_h}
\end{array} \right]
=
v_w \left[ \begin{array}{cc}
\frac{4}{9} & -\frac{17}{12} \\
\frac{41}{9} & -\frac{1}{3}
\end{array} \right] \: .
\eea
We convert Eq.~(\ref{eq:diffm}) into a first order differential equation by
defining $\Psi \equiv \left[ Q \: , \: h \right]^T$, $H \equiv \Psi^\prime$.
We solve the equations seperately for $z<L_w$ (region I), $-L_w < z < 0$
(region II), and $z>0$ (region III) and match $\Psi$ and $H$ across the
boundaries.

In regions I and III, $\gamma_h = 0$, and the solutions are those of the
corresponding homogeneous equation,
\bea
H_{\rm I,III} \left( z \right) = H_H \left( z \right) {\cal D}_{\rm I,III}
& = & Exp \left[ - M^{-1} N z \right] {\cal D}_{\rm I,III} \\
\Psi_{\rm I,III} \left( z \right) & = &
\int_{-\infty}^z dx \; H_H  \left( x \right) {\cal D}_{\rm I,III}
+ {\cal C}_{\rm I,III}
\eea
where ${\cal C}_{\rm I,III}$ and ${\cal D}_{\rm I,III}$
are vectors specifying the boundary
conditions.  The exponential terms grow with $z$, and thus the requirement that
$\Psi (z \rightarrow -\infty) \rightarrow 0$ requires $ {\cal C}_{\rm I}=0$ and
the requirement that $\Psi$ remain finite as $z \rightarrow +\infty$ requires
$ {\cal D}_{\rm III}=0$.

In region II, the solutions will be the sum of a homogeneous solution
with integration constants ${\cal C}_{\rm II}$ and ${\cal D}_{\rm II}$, and a
particular solution,
\bea
H_P (z) & = & - H_H (z)
\int_{-L_w}^z \; dx \: H_H^{-1} (x) \: M^{-1} \:
\left[ \begin{array}{l} \gamma_h \\ 0 \end{array} \right] \: .
\eea
So,
\bea
H_{\rm II} \left( z \right) & = & H_H \left( z \right) {\cal D}_{\rm II}
+ H_P \left( z \right) \; , \\
\Psi_{\rm II} \left( z \right) & = &
\int_{-L_w}^z dx \; \left\{ H_H  \left( x \right) {\cal D}_{\rm II}
+ H_P \left( x \right) \right\}
+ {\cal C}_{\rm II} \; ,
\eea
and matching this to the solutions in regions I and II determines
${\cal D}_{\rm I} = {\cal D}_{\rm II}$, and,
\bea
{\cal D}_{\rm II} & = & \int_{-L_w}^0 dx H_H^{-1} \left( x \right) M^{-1}
\left[ \begin{array}{c} \gamma_h \\ 0 \end{array} \right] \: , \\
{\cal C}_{\rm II} & = & \int_{-\infty}^{-L_w} dx H_H \left( x \right)
{\cal D}_{\rm II} \; , \\
{\cal C}_{\rm III} & = & \int_{-\infty}^0 dx H_H \left( x \right)
{\cal D}_{\rm II} - \int_{-L_w}^0 dx \int_{-L_w}^x dy H_H \left( x
\right) H_H^{-1} \left( y \right) M^{-1} \left[ \begin{array}{c}
\gamma_h \\ 0 \end{array} \right] \; . \eea Note that $\Psi ( z )
= {\cal C}_{\rm III}$ for $z>0$, so this last expression is in
fact the final densities produced by the phase transition.

Since $L_w$ is small, we can derive an approximate form based on
the limit $L_w \rightarrow 0$.  To leading order in $L_w$, we have
\bea
{\cal C}_{\rm III}
= {\cal C}_{\rm II} & = & \int_{-\infty}^{0} dx H_H \left( x
\right) {\cal D}_{\rm II} = - L_w N^{-1} \left[
\begin{array}{c} \gamma_h \\ 0 \end{array} \right] \; , \eea where
\bea {\cal D}_{\rm II} & = & L_w M^{-1} \left[ \begin{array}{c}
\gamma_h \\ 0
\end{array} \right] \: .
\eea
Using the expression for $N^{-1}$, we find the final lepton number
density
\bea
\label{L3}
L_3 = -Q = -\frac{12 L_w \gamma_h}{227 v_w}  ~.
\eea
Recalling equation (\ref{4-14}) that
$\gamma_h \propto v_w/L_w$, we find that our final result is
approximately independent of
the diffusion constants, the bubble wall velocity, and the bubble wall
width, as long as $\Gamma_1$ is fast enough.

\begin{figure}[tbp]
  \begin{center}
    \psfig{figure=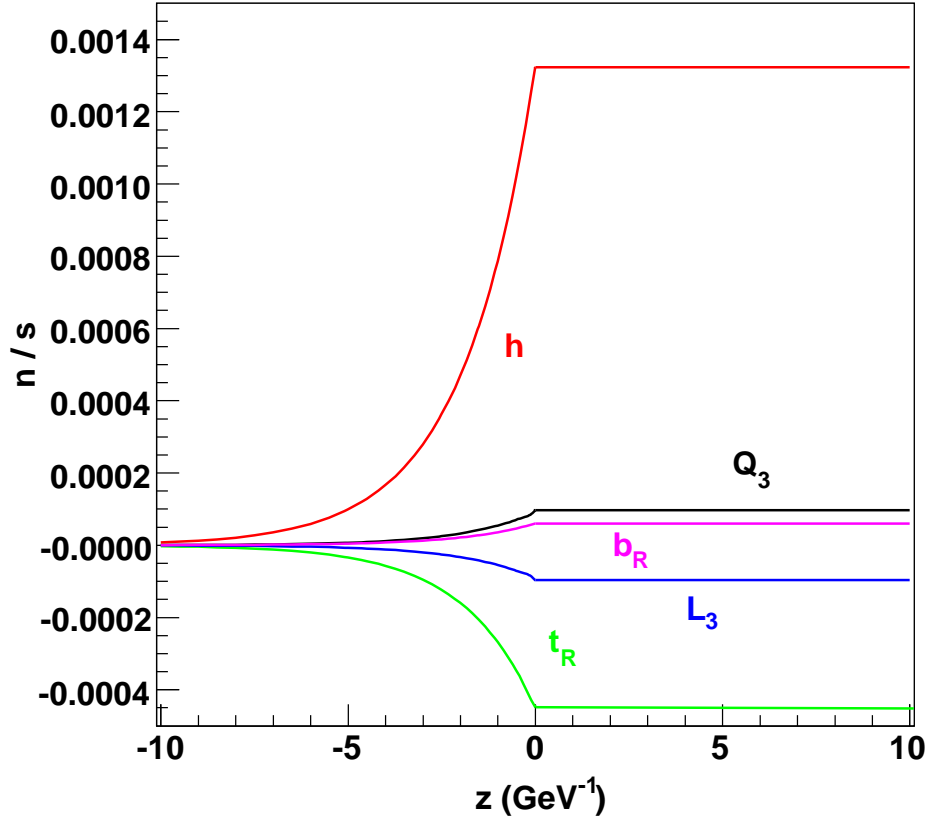,height=12cm}
  \caption{Particle number densities normalized to entropy as a function of
spatial position $z$ for a bubble whose wall is at $z \sim 0$ and parameters
as described in the text.  From top to bottom, the curves represent
the following densities: $h$, $Q_3$,
$b_R$, $L_3$, and $t_R$.}
  \label{fig:densities}
  \end{center}
\end{figure}

Assembling the results, in Figure~\ref{fig:densities} we plot the
densities normalized to the entropy, $s \sim 2 \pi^2 / 45 g_*
T^3$, where $g_* \sim 100$. The densities $L_3$, $t$, and $b$ are
determined using the relations
Eqs.~(\ref{eq:const1}-\ref{eq:const2}).  The wall region is
too small ($2 \times 10^{-3}$~GeV$^{-1}$), and is only in the figure
For the chosen parameters, the ratio of the lepton density to the
entropy density at is given by
\begin{equation}
L_3/s \sim 10^{-4},
\label{eq:l3overs}
\end{equation}
which results, after the inclusion of the dilution factor,
Eq.~(\ref{eq:dilution}) in a final
baryon asymmetry after the electroweak phase transition of
approximately
\begin{equation}
\frac{n_B}{s} \simeq 10^{-10},
\end{equation}
exactly as observed.  Of course, the particular value is highly dependent on
our choice of parameters, but the ability of the Topflavor model to produce
this value is not; the fact that the order of magnitude comes out correctly
is indicitive of the fact that for {\em natural}
values of parameters, this model {\em can} produce
an appropriate baryon number.

\begin{figure}[hbtp]
  \centering
    \psfig{figure=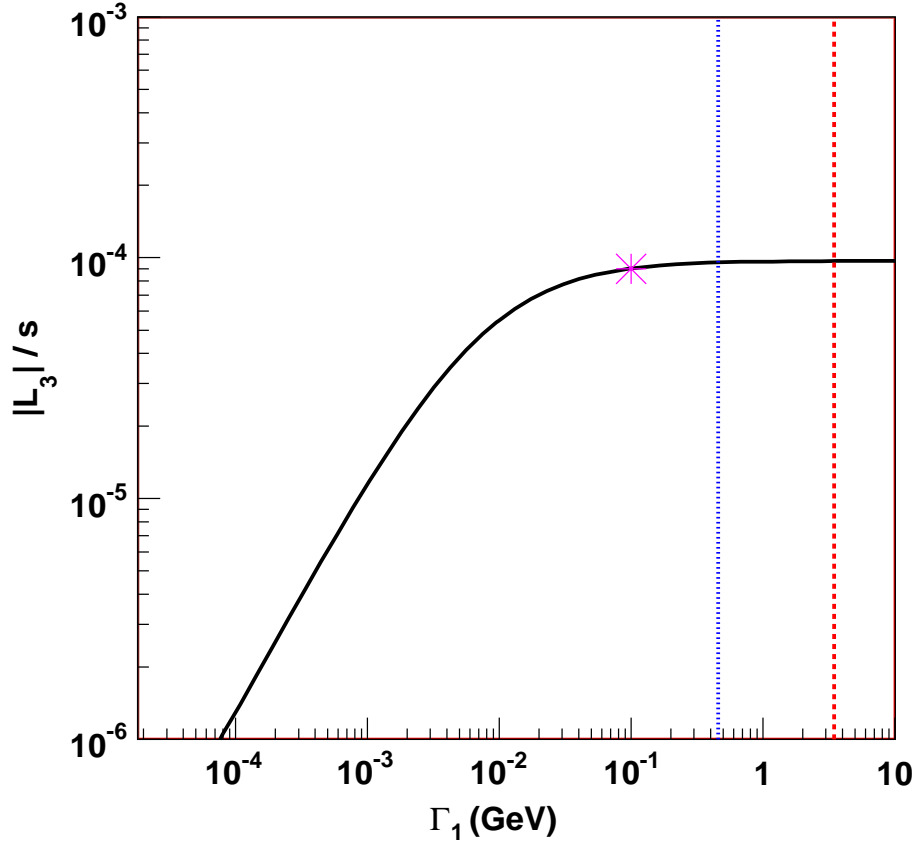,height=12cm}
  \caption{Lepton number density normalized to entropy as a function of
$\Gamma_1$.  The red dashed line is the bound on $\Gamma_1$ from
non-observation of proton decay mediated by instantons, whereas the blue
dotted line is the bound inferred by requiring that the broken phase minimum
remain the true vacuum at $T=0$.  The star indicates the sample
parameter point considered in the text.}
\label{log-lepton-ts-rate}
\end{figure}

{} From our numerical integration of the differential equations, we can
relax our equilibrium assumptions and
examine the effect of finite values for the rates, particularly $\Gamma_1$.
In Figure~ \ref{log-lepton-ts-rate} we present the final lepton density
as a function of $\Gamma_1$ (artificially assuming that
the $CP$-violating source, bubble parameters, and critical temperature are
unchanged).  For small $\Gamma_1$, we see that the final lepton asymmetry
is linearly proportional to $\Gamma_1$.  In this regime, the generated
Higgs/top number densities are
a constant background source, only a small fraction
of which is converted into leptons by $\Gamma_1$.  For large $\Gamma_1$
instead, the equilibrium condition is reached and the dependence on
$\Gamma_1$ saturates.  We see from Figure~\ref{log-lepton-ts-rate}
that our chosen parameters
are just at the turn-on of this saturation region.

\section{Discussion}
\label{discussion}

We have explored the possibility that the baryon asymmetry of the
Universe might have come from new gauge dynamics at the TeV scale.
The primary impediment to realizing this idea is the washout by
ordinary EW sphalerons, which we have avoided by having new
physics which coupled asymmetrically to the three (lepton)
generations\footnote{A parallel idea, which we chose not to
explore in detail, is to have new physics coupled asymmetrically
between quarks and leptons, as in \cite{Georgi:1989xz}.  Anomaly
cancellation is very challenging in this framework, and the
resulting models are therefore somewhat contrived.}. In
particular, we showed that the Topflavor model, with $SU(2)_1
\times SU(2)_2$ breaking at the TeV scale, can easily produce an
appropriate baryon asymmetry for natural values of its parameters.

It is interesting to compare the results obtained in
Eq.~(\ref{eq:l3overs}) with the much smaller ones that are
obtained by similar methods in the MSSM~\cite{CQRVW}. The
difference stems from three main different factors: First, the
magnitude of $\Gamma_1$ is much larger than the EW sphaleron rate
$\Gamma_{\rm ws} \simeq 10^{-4}$~GeV. As shown in
Fig.~\ref{log-lepton-ts-rate}, this induces an enhancement of a
few orders of magnitude in the final result.  Second, the value of
$\Delta \theta \simeq 0.2$ is two orders of magnitude larger than
the typical values of $\Delta \beta \simeq {\cal{O}}\left( 10^{-3}
\right)$  obtained  in the MSSM for moderate values of the CP-odd
Higgs mass. Third, in the MSSM there is a suppression proceeding
from the effective cancellation of chiral charges discussed in
Refs.~\cite{GS, Huet:1995sh} that is not present in the model
analyzed in this article.

We expect that the LHC can extensively test this idea.  First, by
discovering the SM Higgs boson and searching for the existence of
other elements strongly coupled to it, the LHC can help construct
the picture of the ordinary EW phase transition as first or second
order.  This would, for example, confirm or rule out the MSSM
scenario of EW baryogenesis.  Second, by discovering new elements,
in the case of Topflavor, the heavy $W^\prime$ and $Z^\prime$
bosons with masses of about 2 TeV and preferred decays into
third family fermions, should be easily discovered at the LHC
\cite{Sullivan:2003xy}. Finally, by determining the properties of
the new elements, one may determine whether or not they are viable
as a mechanism of baryogenesis.  In the case of Topflavor, the
need for a first order phase transition in the breakdown of
$SU(2)_1 \times SU(2)_2$ is connected to both a small quartic for
the Higgs $\sigma$ and a reasonably strong gauge coupling $g_1$.
The first of these has the dramatic consequence that the mass of
$\sigma$ is well below the symmetry-breaking scale, $m_{\sigma}^2
\sim \lambda u^2$.  For the parameters we explore in detail,
$m_\sigma \sim 600$ GeV, making $\sigma$ the lightest of the new
particles. It couples strongly to the Higgs sector, and to the new
gauge bosons.

An extended gauge sector at the TeV scale has motivations from
many mysteries in particle physics.  In addition, it allows a much
less constrained structure to explore the idea of baryogenesis
through gauge symmetry breaking.  The SM picture arising from the
EW breakdown is economical, but is under assault from the null LEP
Higgs searches, favoring a second order phase transition, and from
the lack of sufficient $CP$ violation to produce enough asymmetry.
Alternatives based on the ordinary EW transition such as in the
MSSM can be viable, but remain constrained by the Higgs mass, and
from the fact that the new sources of $CP$ violation remain
tightly connected to low energy $CP$-violating observables.
Extended gauge symmetries can naturally have first order
transitions, and $CP$ violation is further removed from low energy
observables.  The primary obstacle is the fact that the EW
sphalerons will try to erase any generated baryon asymmetry with
$B-L=0$, but this is overcome if the SM matter does not couple
universally to the new dynamics. Ultimately, the LHC will help in
resolving this question.

\section{ACKNOWLEDGMENTS}
We would like to thank Junhai Kang, and David Morrisey for helpful
conversations and Ignacy Sawicki for computing assistance. Work at
ANL is supported in part by the US DOE, Div.\ of HEP, Contract
W-31-109-ENG-38. J.S. would like to acknowledge Prospects in
Theoretical Physics 05 for part of the initial work was done
there, T.T. and C.W. would like to thank the Aspen Center for
Physics were an important part of this work has been performed.


\parindent 0pt
\begin{thebibliography}{40}

\bibitem{Sak67} A. D. Sakharov, Pis'ma Zh. Eksp. Teor. Fiz. {\bf 5}, 32 (1967).

\bibitem{Yoshimura}
M.~Yoshimura, Phys.\ Rev.\ Lett.\  {\bf 41}, 281 (1978)
[Erratum-ibid.\  {\bf 42}, 746 (1979)].

\bibitem{SDLS}
S.~Dimopoulos and L.~Susskind, Phys.\ Rev.\ D {\bf 18}, 4500
(1978).

\bibitem{Weinberg}
S.~Weinberg, Phys.\ Rev.\ Lett.\  {\bf 42}, 850 (1979).

\bibitem{EKMST}
E. W. Kolb and M. S. Turner, {\it The Early Universe},
Addison-Wesley Publishing Company (1990).

\bibitem{LG86}
M. Fukugita and T. Yanagida, Phys. Lett. B {\bf 174}, 45 (1986).

\bibitem{WBMP}
W.~Buchmuller and M.~Plumacher, Int.\ J.\ Mod.\ Phys.\ A {\bf 15},
5047 (2000), and references therein.

\bibitem{IAMD}
I.~Affleck and M.~Dine, Nucl.\ Phys.\ B {\bf 249}, 361 (1985).

\bibitem{Kuzmin:1985mm}
  V.~A.~Kuzmin, V.~A.~Rubakov and M.~E.~Shaposhnikov,
  Phys.\ Lett.\ B {\bf 155}, 36 (1985).

\bibitem{BG} For reviews, see, M.~Trodden,
  Rev.\ Mod.\ Phys.\  {\bf 71}, 1463 (1999); V.A. Rubakov, M. E. Shaposhnikov, Usp. Fiz.Nauk \textbf{166}
493 (1996); Phys. Usp. \textbf{39} 461 (1996); A. G. Cohen, D. B.
Kaplan, A. E. Nelson, Ann. Rev. Nucl. Part. Sci. \textbf{43} 27
(1993).

\bibitem{Barate:2003sz}
  R.~Barate {\it et al.}  [LEP Working Group for Higgs boson searches],
  Phys.\ Lett.\ B {\bf 565}, 61 (2003).

\bibitem{Pietroni}
M.~Pietroni,
Nucl. Phys. B {\bf 402}, 27 (1993).

\bibitem{ADFM}
 A.~T.~Davies, C.~D.~Froggatt and R.~G.~Moorhouse,
Phys.\ Lett.\ B {\bf 372}, 88 (1996).

\bibitem{SJHMGS}
  S.~J.~Huber and M.~G.~Schmidt,
  Nucl.\ Phys.\ B {\bf 606}, 183 (2001).

\bibitem{Kang:2004pp}
  J.~Kang, P.~Langacker, T.~j.~Li and T.~Liu,
  Phys.\ Rev.\ Lett.\  {\bf 94}, 061801 (2005).

\bibitem{Menon:2004wv}
  A.~Menon, D.~E.~Morrissey and C.~E.~M.~Wagner,
  Phys.\ Rev.\ D {\bf 70}, 035005 (2004).

\bibitem{CJK}
J.~M.~Cline, M.~Joyce and K.~Kainulainen,
JHEP {\bf 0007}, 018 (2000).

\bibitem{CK}
J.~M.~Cline and K.~Kainulainen,
Phys.\ Rev.\ Lett.\  {\bf 85}, 5519 (2000).

\bibitem{CQRVW}
  M.~Carena, M.~Quiros, A.~Riotto, I.~Vilja and C.~E.~M.~Wagner,
  Nucl.\ Phys.\ B {\bf 503}, 387 (1997).

\bibitem{MCQCW}
M.~Carena, M.~Quiros and C.~E.~M.~Wagner, Nucl.\ Phys.\ B {\bf
524}, 3 (1998).

\bibitem{CMQSW1}
M.~Carena, J.~M.~Moreno, M.~Quiros, M.~Seco and C.~E.~M.~Wagner,
Nucl.\ Phys.\ B {\bf 599}, 158 (2001).

\bibitem{CQSW2}
M.~Carena, M.~Quiros, M.~Seco and C.~E.~M.~Wagner, Nucl.\ Phys.\ B
{\bf 650}, 24 (2003).

\bibitem{KPSW1}
K.~Kainulainen, T.~Prokopec, M.~G.~Schmidt and S.~Weinstock, JHEP
{\bf 0106}, 031 (2001).

\bibitem{KPSW2}
K.~Kainulainen, T.~Prokopec, M.~G.~Schmidt and S.~Weinstock,
Phys.\ Rev.\ D {\bf 66}, 043502 (2002).

\bibitem{Prokopec:2003pj}
  T.~Prokopec, M.~G.~Schmidt and S.~Weinstock,
  Annals Phys.\  {\bf 314}, 208 (2004)
  [arXiv:hep-ph/0312110];
  T.~Prokopec, M.~G.~Schmidt and S.~Weinstock,
  Annals Phys.\  {\bf 314}, 267 (2004)
  [arXiv:hep-ph/0406140].

\bibitem{Konstandin:2005cd}
  T.~Konstandin, T.~Prokopec, M.~G.~Schmidt and M.~Seco,
  Nucl.\ Phys.\ B {\bf 738}, 1 (2006)
  [arXiv:hep-ph/0505103].

\bibitem{Joyce:1996cp}
M.~Joyce,
Phys.\ Rev.\ D {\bf 55}, 1875 (1997).
T.~Prokopec,
Phys.\ Lett.\ B {\bf 483}, 1 (2000).


\bibitem{Joyce:1997fc}
M.~Joyce and T.~Prokopec,
Phys.\ Rev.\ D {\bf 57}, 6022 (1998).

\bibitem{Davidson:2000dw}
S.~Davidson, M.~Losada and A.~Riotto,
Phys.\ Rev.\ Lett.\  {\bf 84}, 4284 (2000).

\bibitem{Servant:2001jh}
  G.~Servant,
  JHEP {\bf 0201}, 044 (2002).

\bibitem{Carena:2004ha}
  M.~Carena, A.~Megevand, M.~Quiros and C.~E.~M.~Wagner,
  Nucl.\ Phys.\ B {\bf 716}, 319 (2005).

\bibitem{Provenza:2005nq}
  A.~Provenza, M.~Quiros and P.~Ullio,
  JHEP {\bf 0510}, 048 (2005).

\bibitem{Kuzmin:1987wn}
  V.~A.~Kuzmin, V.~A.~Rubakov and M.~E.~Shaposhnikov,
  Phys.\ Lett.\ B {\bf 191}, 171 (1987).

\bibitem{Dreiner:1992vm}
  H.~K.~Dreiner and G.~G.~Ross,
  Nucl.\ Phys.\ B {\bf 410}, 188 (1993).

\bibitem{Chivukula:1995gu}
  R.~S.~Chivukula, E.~H.~Simmons and J.~Terning,
  Phys.\ Rev.\ D {\bf 53}, 5258 (1996);
  E.~Malkawi, T.~Tait and C.~P.~Yuan,
  Phys.\ Lett.\ B {\bf 385}, 304 (1996);
  D.~J.~Muller and S.~Nandi,
  Phys.\ Lett.\ B {\bf 383}, 345 (1996);
  P.~Batra, A.~Delgado, D.~E.~Kaplan and T.~M.~P.~Tait,
  JHEP {\bf 0402}, 043 (2004).

\bibitem{Morrissey:2005uz}
  D.~E.~Morrissey, T.~M.~P.~Tait and C.~E.~M.~Wagner,
  Phys.\ Rev.\ D {\bf 72}, 095003 (2005).

\bibitem{Huet:1995sh}
  P.~Huet and A.~E.~Nelson,
  Phys.\ Rev.\ D {\bf 53}, 4578 (1996).

\bibitem{Col78}
S. Coleman, V. Glaser and A. Martin, {\it Commun. Math. Phys. }
{\bf 58} 211 (1978).
%

\bibitem{Moore:1995si}
  G.~D.~Moore and T.~Prokopec,
  Phys.\ Rev.\ D {\bf 52}, 7182 (1995)
  [arXiv:hep-ph/9506475];
  G.~D.~Moore and T.~Prokopec,
  Phys.\ Rev.\ Lett.\  {\bf 75}, 777 (1995)
  [arXiv:hep-ph/9503296].

\bibitem{Kosowsky:1992rz}
  A.~Kosowsky, M.~S.~Turner and R.~Watkins,
  Phys.\ Rev.\ Lett.\  {\bf 69}, 2026 (1992).

\bibitem{Apreda:2001us}
  R.~Apreda, M.~Maggiore, A.~Nicolis and A.~Riotto,
  Nucl.\ Phys.\ B {\bf 631}, 342 (2002).

\bibitem{Nicolis:2003tg}
  A.~Nicolis,
  Class.\ Quant.\ Grav.\  {\bf 21}, L27 (2004).

\bibitem{Grojean:2006bp}
  C.~Grojean and G.~Servant,
  [arXiv:hep-ph/0607107].

\bibitem{Randall:2006py}
  L.~Randall and G.~Servant,
  [arXiv:hep-ph/0607158].

\bibitem{STW}
  J.~Shu, T.~M.~P.~Tait and C.~E.~M.~Wagner, in
  progress.


\bibitem{Linde:1981zj}
  A.~D.~Linde,
  Nucl.\ Phys.\ B {\bf 216}, 421 (1983)
  [Erratum-ibid.\ B {\bf 223}, 544 (1983)].


\bibitem{Dine:1992wr}
  M.~Dine, R.~G.~Leigh, P.~Y.~Huet, A.~D.~Linde and D.~A.~Linde,
  Phys.\ Rev.\ D {\bf 46}, 550 (1992).

\bibitem{Cohen:1994ss}
  A.~G.~Cohen, D.~B.~Kaplan and A.~E.~Nelson,
  Phys.\ Lett.\ B {\bf 336}, 41 (1994).

\bibitem{Joyce:1994zn}
  M.~Joyce, T.~Prokopec and N.~Turok,
  %
  Phys.\ Rev.\ D {\bf 53}, 2930 (1996).

\bibitem{Arnold:1996dy}
  P.~Arnold, D.~Son and L.~G.~Yaffe,
  Phys.\ Rev.\ D {\bf 55}, 6264 (1997);
  D.~Bodeker, G.~D.~Moore and K.~Rummukainen,
  Nucl.\ Phys.\ Proc.\ Suppl.\  {\bf 83}, 583 (2000);
  G.~D.~Moore,
  [arXiv:hep-ph/0009161].



\bibitem{Riotto:1995hh}
  A.~Riotto,
  Phys.\ Rev.\ D {\bf 53}, 5834 (1996).

\bibitem{Riotto:1997vy}
  A.~Riotto,
  Nucl.\ Phys.\ B {\bf 518}, 339 (1998).

\bibitem{Georgi:1989xz}
  H.~Georgi, E.~Jenkins and E.~H.~Simmons,
  Nucl.\ Phys.\ B {\bf 331}, 541 (1990).

\bibitem{GS}
  G.~F.~Giudice and M.~E.~Shaposhnikov,
  Phys.\ Lett.\ B {\bf 326}, 118 (1994).

\bibitem{Sullivan:2003xy}
  Z.~Sullivan,
  [arXiv:hep-ph/0306266].




















\end{thebibliography}
\end{document}